# Three-Dimensional Phase Field Simulations of Hysteresis and Butterfly Loops by Finite Volume Method*

MA  Zhi(马治)**,  XI  Li-Ying(席丽莹),  CHEN  Huan-Ming(陈焕铭), ZHENG Fu(郑富), GAO Hua(高华), TONG Yang(童洋)

*School of Physics & Electrical Information Engineering, Ningxia University, Yinchuan 750021, PR China*

*Supported by the Research Starting Funds for Imported Talents of Ningxia University (Grant No. BQD2012011).

**Corresponding author. Email address: mazhicn@126.com (M. Z.);

Tel: 0951-2061430        Fax: 0951-2061430

Three-dimensional (3D) simulations of ferroelectric hysteresis and butterfly loops have been carried out based on solving the time dependent Ginsburg-Landau equations by using a finite volume method. The influence of externally mechanical loadings with a tensile strain and a compressive strain on the hysteresis and butterfly loops has been studied numerically. The 3D ferroelectric domain formation and its evolution have also been presented in the paper. The simulation results successfully reveal the macroscopically nonlinear response to the applied stresses and electric field.

Ferroelectricity is one of the fastest growing fields during the past several decades. Interest in this field is attributable to the increasing numbers of practical applications in micro-electromechanical systems, microwave devices, and memory devices [1-4]. It could be used as pressure sensors, ultrasonic motors, transducers and varied actuators [5-8]. However, the constitutive relation of ferroelectric materials possesses so strong and complex nonlinear properties that it is difficult to accurately model the hysteresis and butterfly loops.

Thermodynamic phenomenological theory is very convenient and more accurate to describe structural phase transitions compared to other theories, such as first principles calculations or Ab initio molecular dynamics simulations. Landau theory has played an important role in understanding the thermodynamics of ferroelectric phase transitions [9]. If the free energy of ferroelectric system can be constructed accurately, the phenomenological theory can even be used to describe complicated phase transitions that involve the change of unit cell [10].

Recently there have been a number of two dimensional computer simulations of domain structure evolution during ferroelectric transitions using the time-dependent Ginsburg Landau (TDGL) equation field model [11]. However, all of these previous studies were performed in two dimensions and using the same method: A complicated semi-implicit Fourier-spectral method [12-15]. Undoubtedly a more realistic 3D simulation and a more simple method are desirable. In this paper, a three-dimensional simulation of ferroelectric hysteresis and butterfly loops have been carried out based on solving the time dependent Ginsburg-Landau equations by using a finite volume method (FVM). The influence of externally mechanical loadings with a tensile strain and a compressive strain on the hysteresis and butterfly loops has also been investigated.

Ferroelectric material possesses spontaneous polarization, spontaneous strain and a domain structure below its Curie-Weiss temperature. In phase field simulations, the temporal evolution of the domain structure is governed by the time-dependent Ginzburg-Landau equation as following [16]:

$$\frac{\partial P_i(\vec{r},t)}{\partial t} = -L \frac{\delta F}{\delta P_i(\vec{r},t)} \qquad (i=1, 2, 3) \qquad (1)$$

where $P_i$ is used as the order parameters, $L$ denotes the kinetic coefficient. $F$ is the total free energy of the system, $\delta F / \delta P_i(\vec{r},t)$ represents the thermodynamic driving force for the spatial and temporal evolution of the simulated system. In the present study, we adopt the above equation to simulate the spontaneous polarization in a ferroelectric under an applied electrical or mechanical field. The total free energy includes the bulk free energy, the domain wall energy or the polarization gradient energy, the elastic strain energy and the electric energy. The total free energy of the system is given by:

$$F = \int [f_L + f_G + f_E + f_{elec}] dV \qquad (2)$$

We employ a six-order polynomial for the Landau bulk-free energy density [17]:

$$\begin{aligned} f_L(P_i) &= \alpha_1(P_1^2 + P_2^2) + \alpha_3 P_3^2 + \alpha_{11}(P_1^4 + P_2^4) \\ &+ \alpha_{33} P_3^4 + \alpha_{13}(P_1^2 P_3^2 + P_2^2 P_3^2) + \alpha_{12} P_1^2 P_2^2 \\ &+ \alpha_{111}(P_1^6 + P_2^6) + \alpha_{333} P_3^6 \end{aligned} \qquad (3)$$

where $\alpha_1$, $\alpha_3$, $\alpha_{11}$, $\alpha_{33}$, $\alpha_{13}$, $\alpha_{12}$, $\alpha_{111}$, $\alpha_{333}$ are constant coefficients. For simplicity, the lowest order of the gradient energy density is used here, which takes the form as follows:

$$f_G(P_{i,j}) = \frac{1}{2} G_{ijkl} P_{i,j} P_{k,l}$$
$$= \frac{1}{2} G_{11}(P_{1,1}^2 + P_{1,2}^2 + P_{1,3}^2 + P_{2,1}^2 + P_{2,2}^2 + P_{2,3}^2 + P_{3,1}^2 + P_{3,2}^2 + P_{3,3}^2) \quad (4)$$

where $G_{ij}$ denotes the gradient coefficients. The comma in the subscripts denotes spatial differentiation. When an external electric field $E_i$ is applied, the additional electrical energy density is given by:

$$f_{elec}(P_i, E_i) = -E_i P_i = -(E_1 P_1 + E_2 P_2 + E_3 P_3) \quad (5)$$

Regarding the symmetry of the polarized state, the extension of the elastic energy density takes the following form:

$$f_E(P_i, \varepsilon_{ij}) = \frac{1}{2} c_{ijkl} (\varepsilon_{ij} - \varepsilon_{ij}^0)(\varepsilon_{kl} - \varepsilon_{kl}^0) \quad (6)$$

where $c_{ijkl}$ is the elastic stiffness tensor, $\varepsilon_{ij}$ is the total strain, $\varepsilon_{ij}^0$ is the electrostrictive strain caused by the polarization field and it could be expressed as $\varepsilon_{ij}^0 = Q_{ijkl} P_k P_l$. The spontaneous strains are linked to the spontaneous polarization components in ferroelectric material in the following form:

$$\varepsilon_{11}^0 = Q_{11} P_1^2 + Q_{12} P_2^2 + Q_{13} P_3^2 \quad (7)$$

$$\varepsilon_{22}^0 = Q_{12} P_1^2 + Q_{11} P_2^2 + Q_{13} P_3^2 \quad (8)$$

$$\varepsilon_{33}^0 = Q_{13} P_1^2 + Q_{13} P_2^2 + Q_{33} P_3^2 \quad (9)$$

$$\varepsilon_{12}^0 = \varepsilon_{21}^0 = Q_{44} P_1 P_2 \quad (10)$$

$$\varepsilon_{23}^0 = \varepsilon_{32}^0 = Q_{44} P_2 P_3 \quad (11)$$

$$\varepsilon_{13}^0 = \varepsilon_{31}^0 = Q_{66} P_1 P_3 \quad (12)$$

where $Q_{ij}$ stants for the electrostrictive constants. The elastic strain energy density can be divided into three parts: the pure elastic energy density, the energy density from the pure polarization and the electrostrictive coupling energy density:

$$f_{E1}(\varepsilon_{ij}) = \frac{1}{2}C_{11}(\varepsilon_{11}^2+\varepsilon_{22}^2) + 2C_{44}(\varepsilon_{23}^2+\varepsilon_{13}^2) + 2C_{66}\varepsilon_{12}^2$$
$$+ \frac{1}{2}C_{33}\varepsilon_{33}^2 + C_{12}\varepsilon_{11}\varepsilon_{22} + C_{13}(\varepsilon_{11}\varepsilon_{33} + \varepsilon_{22}\varepsilon_{33}) \quad (13)$$

$$f_{E2}(P_i) = \beta_{11}(P_1^4 + P_2^4) + \beta_{33}P_3^4 + \beta_{12}P_1^2P_2^2 + \beta_{13}P_1^2P_3^2 + \beta_{23}P_2^2P_3^2 \quad (14)$$

$$f_{E3}(\varepsilon_{ij}, P_i) = -(q_{11}\varepsilon_{11} + q_{12}\varepsilon_{22} + q_{13}\varepsilon_{33})P_1^2$$
$$- (q_{21}\varepsilon_{11} + q_{22}\varepsilon_{22} + q_{33}\varepsilon_{33})P_2^2$$
$$- (q_{31}\varepsilon_{11} + q_{32}\varepsilon_{22} + q_{33}\varepsilon_{33})P_3^2 \quad (15)$$
$$- q_{44}\varepsilon_{12}P_2P_3 - q_{55}\varepsilon_{13}P_1P_3 - q_{66}\varepsilon_{23}P_1P_2$$

where $C_{ij}$ in the energy function should be the components of the elastic stiffness tensor, the coefficients $\beta_{ij}$ are called electricstrictive constants for constant stress. The coefficients $\beta_{ij}$ and $q_{ij}$ could be written as:

$$\beta_{11} = \frac{1}{2}C_{11}(Q_{11}^2 + Q_{12}^2) + \frac{1}{2}C_{33}Q_{13}^2 + C_{12}Q_{11}Q_{12} + C_{13}Q_{11}Q_{13} + C_{13}Q_{12}Q_{13} \quad (16)$$

$$\beta_{33} = C_{11}Q_{13}^2 + C_{12}Q_{13}^2 + \frac{1}{2}C_{33}Q_{33}^2 + 2C_{13}Q_{13}Q_{33} \quad (17)$$

$$\beta_{12} = C_{12}Q_{11}^2 + 2C_{11}Q_{11}Q_{12} + C_{12}Q_{12}^2 + 2C_{13}Q_{11}Q_{13} + 2C_{13}Q_{12}Q_{13} + C_{33}Q_{13}^2 + 2C_{44}Q_{66}^2 \quad (18)$$

$$\beta_{13} = C_{11}Q_{11}Q_{13} + C_{12}Q_{11}Q_{13} + C_{11}Q_{12}Q_{13} + C_{12}Q_{12}Q_{13} + 2C_{13}Q_{13}^2 + C_{33}Q_{13}^2 + 2C_{66}Q_{44}^2 \quad (19)$$

$$\beta_{23} = C_{11}Q_{11}Q_{13} + C_{12}Q_{11}Q_{13} + C_{11}Q_{12}Q_{13} + C_{12}Q_{12}Q_{13} + 2C_{13}Q_{13}^2 + C_{33}Q_{13}^2 + 2C_{44}Q_{66}^2 \quad (20)$$

$$q_{11} = q_{22} = C_{11}Q_{11} + C_{12}Q_{12} + C_{13}Q_{13} \quad (21)$$

$$q_{12} = q_{21} = C_{12}Q_{11} + C_{11}Q_{12} + C_{13}Q_{13} \quad (22)$$

$$q_{13} = q_{23} = C_{13}Q_{11} + C_{13}Q_{12} + C_{13}Q_{13} \quad (23)$$

$$q_{31} = q_{32} = C_{13}Q_{13} + C_{12}Q_{13} + C_{13}Q_{33} \quad (24)$$

$$q_{33} = 2C_{13}Q_{13} + C_{33}Q_{33} \quad (25)$$

$$q_{44} = 4C_{44}Q_{44} \quad (26)$$

$$q_{55} = 4C_{66}Q_{44} \quad (27)$$

$$q_{66} = 4C_{44}Q_{66} \quad (28)$$

For convenience of simulations, the set of normalized and dimensionless variables, as described in the literature [14], is employed in the present simulation.

In the present study, the TDGL equations are solved by using cell centered finite volume method [18]. Finite volume method requires to divide domain of our problem into non-overlapping cells and is capable of numerical solving of the canonical governing equation like

$$\frac{\partial(\rho\psi)}{\partial t} = \nabla \cdot (\vec{u}\psi) + [\nabla \cdot (\Gamma_i \nabla)]^n \psi + S_\psi \qquad (29)$$

where $n$ is integer and $\rho$, $\vec{u}$ and $\Gamma_i$ can be arbitrary functions of any variable of the system. Lead titanate is chosen to carry out the simulation. Mayavi is employed to show the 3D evolution patterns [19]. A 3D simulation of ferroelectrics subjected to electric-mechanical loading is carried out. A discrete grid of 50×50×10 points with a cell size of $\Delta x^* = \Delta y^* = \Delta z^* = 0.25$ and fixed value boundary conditions are implemented in left-right, front-back and top-bottom faces. As the finite volume method is used to ensure high accuracy even in the case of large cells, the time step is set as 0.01. Values of the normalized coefficients used in the simulation are listed in table 1.

Figure1 shows the influence of mechanical loading on the hysteresis loops. Two normalized external mechanical loading are applied parallel to the electric field. It can be seen from the loops that the applied tensile loading leads to an increase of the polarization and the coercive field. The applied compressive loading will leads to a decrease of the polarization and the coercive field. The influence of stress on the coercive electric field can be explained by the switching criterion [20, 21]. According to the theory, the external applied loading affect the mechanical work in the switching criterion, while the total work required for polarization switching is constant. Therefore, the electric work and the coercive field will need to adjust its magnitude.

According to Landau-Devonshire theory [22], the remanent strain of a ferroelectric film is approximately proportional to the square of the remanent polarization, it could be written as $\bar{\varepsilon}_{33}^r = Q_{13}\overline{P}_3^r$, where $Q_{13}$ is the electrostriction coefficient. The influences of compressive and tensile mechanical loading on the butterfly and hysteresis loops are simulated. Furthermore, the case with no mechanical loading is also simulated for comparison. The normalized electric field along the z direction is applied. Figure 2 shows the influence of mechanical loading on (a) hysteresis loops and (b) butterfly loops. It was found that the influences of compressive and tensile mechanical loading on the hysteresis loop are similar to figure 1 result, but the butterfly loop shows an asymmetric style. The asymmetric behavior could also be observed in other simulated works [14, 23]. It might be attributed to the insufficient number of domains used in the present study.

Temporal evolution of morphological patterns without mechanical loading is shown in figure 3. The evolution starts from a noised field, which is represented by a polarization vector field in which each unit is in random direction, and it has small magnitude. Figure 3 illustrates the polarization after 50, 100, 200, 300, 400, 500, 600 and 1000 time steps of evolution. It can be seen that the polarization begins to grow and the rudiment of the domain structure starts to appear. After 50 time steps, the polarization is still in nucleation state. After about 100 time steps, the nucleation stage is completed and all four possible domain orientations are present. The evolution of the simulated domain structure shows that the formation of a stable domain experienced three classical processes, i.e., nucleation, growing accompanied with wall moving and the stable stage. These results are consistent with those of the simulations [24, 25].

In conclusion, a finite volume method was proposed to simulate the hysteresis and butterfly loops for 3D ferroelectrics subjected to various constant mechanical loadings coupled with electric loading. Real-space phase field model was employed to simulate the evolution of ferroelectric domains. It is demonstrated that this method is useful to analyze the influence of mechanical loading on the hysteresis and butterfly loops. Based on this method, it is easier to obtain the 3D temporal evolution patterns with/without compressive and tensile mechanical loadings. The related electromechanical behaviors induced under different inducing conditions were also analyzed numerically. The results in this work are helpful for investigating the macroscopic response of ferroelectrics subjected to external applied electric and mechanical loadings and can give insight into the understanding of domain structures and the macroscopic ferroelectrics.

**Figure Caption**

Fig. 1. Simulated hysteresis loops of average polarization versus applied electric field.

Fig. 2. Influence of mechanical loading on (a) hysteresis loops and (b) butterfly loops.

Fig. 3. Temporal evolution of the morphological patterns without mechanical loading

**Table Caption**

**Table 1** Normalized materials coefficients used in the simulation

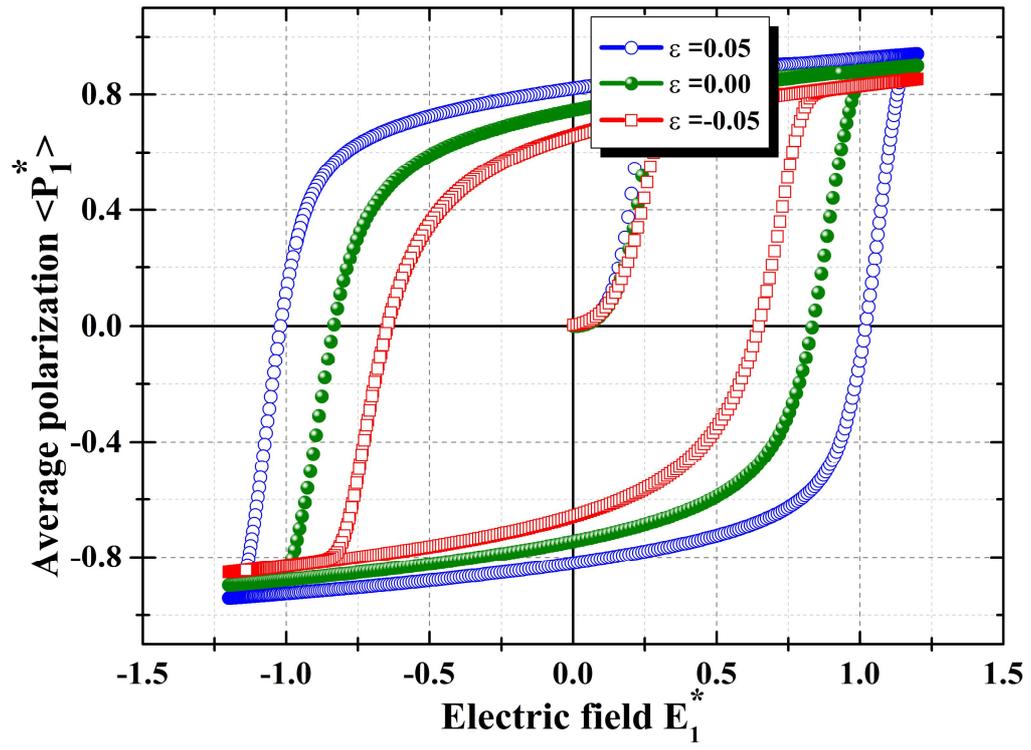

**Fig. 1**.

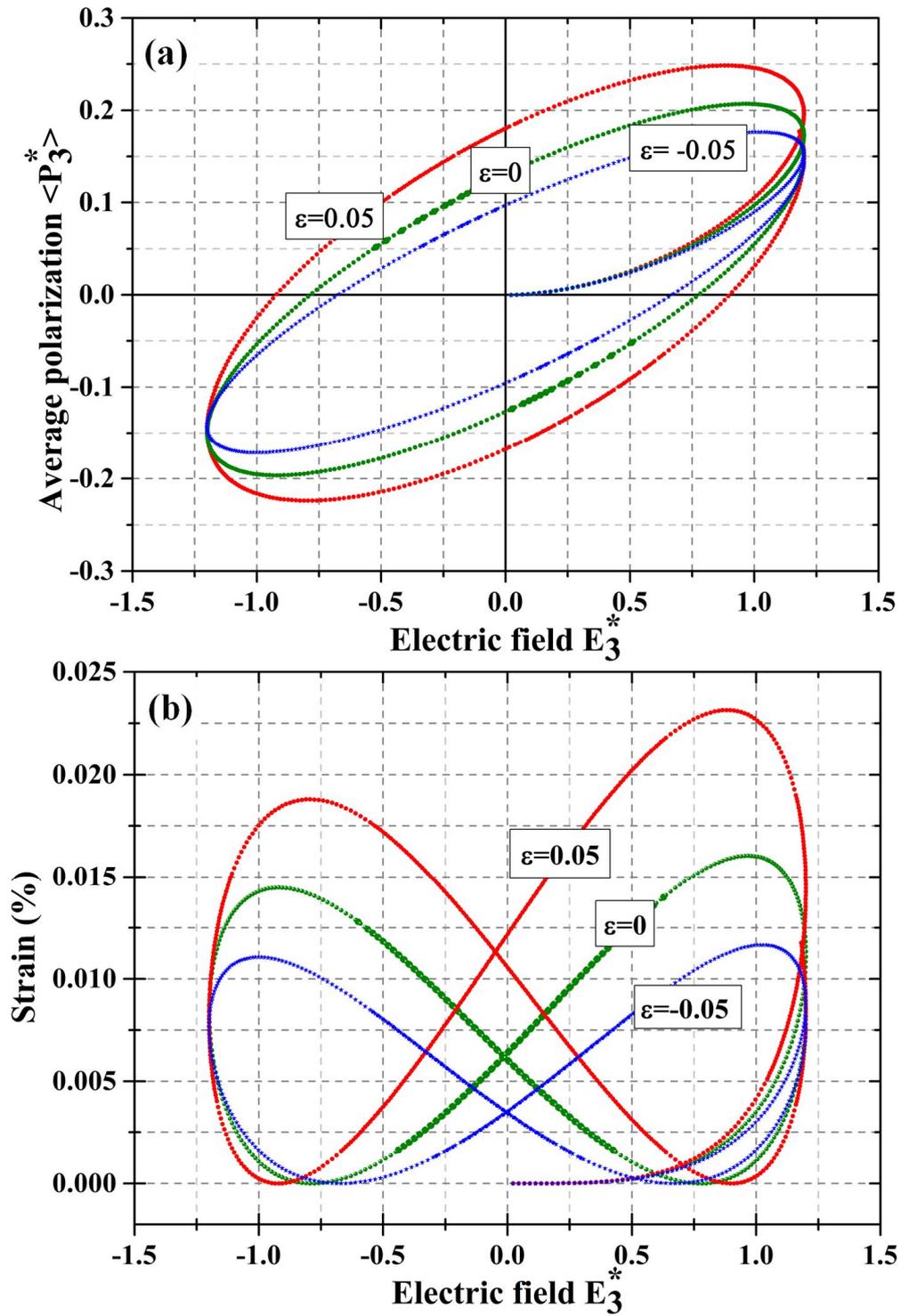

**Fig. 2**.

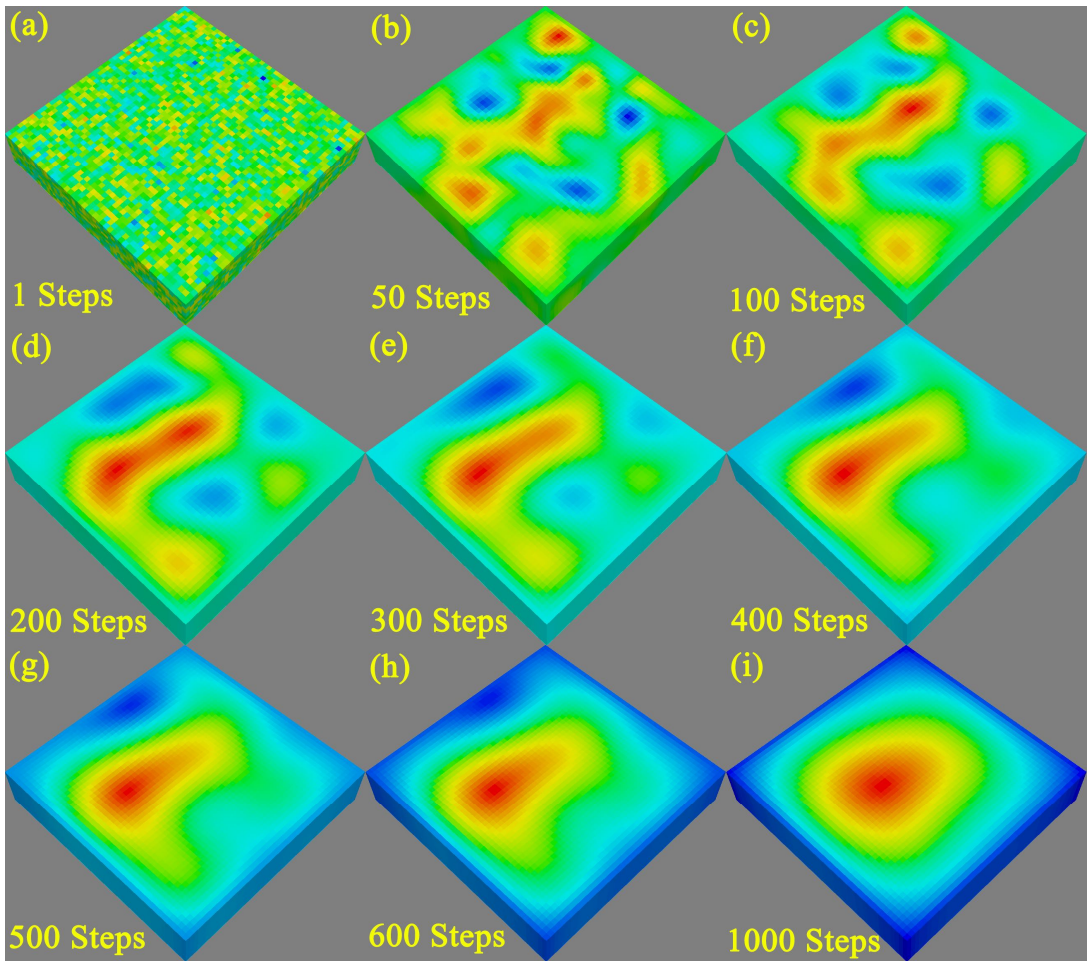

**Fig. 3**.

**Table 1**

| Landau coefficient | Elastic constant | Electrostrictive constant | Gradient coefficient |
|---|---|---|---|
| $\alpha_1^* = -1.00$ | $c_{11}^* = 2395$ | $Q_{11}^* = 0.00370$ | $G_{11}^* = -1.0$ |
| $\alpha_3^* = 1.90$ | $c_{12}^* = 520$ | $Q_{12}^* = -0.00313$ | |
| $\alpha_{11}^* = -0.25$ | $c_{13}^* = 476$ | $Q_{13}^* = 0.00375$ | |
| $\alpha_{12}^* = 0.32$ | $c_{33}^* = 533$ | $Q_{33}^* = 0.00375$ | |
| $\alpha_{13}^* = 0.36$ | $c_{44}^* = 476$ | $Q_{44}^* = 0.00775$ | |
| $\alpha_{23}^* = 0.38$ | $c_{66}^* = 552$ | $Q_{66}^* = 0.00875$ | |
| $\alpha_{33}^* = 0.40$ | | | |
| $\alpha_{111}^* = 0.57$ | | | |
| $\alpha_{333}^* = 0.58$ | | | |